# Pyroptosis: Physiological roles in viral infection

Nelson Durán[1,2] and Wagner J. Fávaro[1]

[1]Laboratory of Urogenital Carcinogenesis and Immunotherapy, Department of Structural and Functional Biology, University of Campinas, Campinas, SP, Brazil,
[2]Nanomedicine Research Unit (Nanomed), Federal University of ABC (UFABC), Santo André, SP, Brazil.

[*]**Correspondence**: Prof. Nelson Durán (E-mail: nelsonduran1942@gmail.com) and Prof. Wagner J. Fávaro (E-mail: favarowj@unicamp.br).

**Abstract**

The current review focuses on important aspects of pyroptosis, such as morphological features of inflammasomes, as well as on knowledge about coronavirus infection mechanisms and the association between SARS-CoV-2 infection and cell pyroptosis. The application of immunomodulation therapies to Covid-19 patients is the most promising treatment under investigation nowadays. The possibilities of assistance to help protecting patients from infections are also addressed in the present study.

**Highlights**

- Inflammatory caspases play an important role in innate immunity.
- Pyroptosis eliminates the reproduction site of intracellular pathogens.
- SARS-CoV proteins triggers NLRP3 inflammasome induction.
- SARS-CoV-2 infection and cell pyroptosis.
- Immunomodulation therapies to Covid-19 patients is the most promising treatment.

## 1. Introduction

It is well known that inflammatory caspases play an important role in innate immunity by responding to cytosolic signals and by inducing a twofold response (Jorgensen and Miao, 2015). First, caspase-1 induces the activation and release of pro-inflammatory cytokines, such as interleukin-18 (IL-18) and IL-1β. Second, caspase-11 or caspase-1 can trigger a lytic, programmed cell death known as pyroptosis. Pyroptosis eliminates the reproduction site of intracellular pathogens, such as bacteria or virus, which makes them susceptible to phagocytosis and to be destroyed by secondary phagocytes. However, anomalous systemic activation of pyroptosis *in vivo* may lead to sepsis in some cases. Several pathogens have evolved to prevent or disrupt pyroptosis, a fact that highlights inflammasomes' ability to recognize viral infections. These effects are also observed in many health conditions (Ma et al., 2018; Liang et al., 2020) such as cardiovascular diseases (Zeng et al., 2019) and cancer (Durán and Fávaro, 2020; unpublished results).

## 2. General aspects of inflammation

It is known that respiratory epithelial cell (REC) death caused by influenza virus infection is responsible for triggering immune response to inflammation; however, the exact cell death mechanisms triggered in these cases remain poorly understood. Data available in the literature have indicated that apoptosis is triggered at initial infection stages although the cell death route is displaced to pyroptosis at late infection stages.

It was suggested that the change from apoptosis to pyroptosis is stimulated by the type I interferon (I-IFN)-mediated JAK-STAT signaling pathway, which inhibits apoptosis through the expression of the Bcl-xL anti-apoptotic gene. Moreover, JAK-STAT signaling suppression has inhibited pyroptosis and increased apoptosis in infected AAH (human atypical adenomatous hyperplasia) and PL16T (respiratory epithelial cells). Accordingly, Lee et al. (2018) have suggested that the type I interferon (I-IFN) signaling route activates pyroptosis, rather than apoptosis, in RECs in a reciprocally exclusive mode and triggers pro-inflammatory response to influenza virus infection (Lee et al., 2018).

The important role played by inflammatory response induction has been emphasized by several studies, so far. The inflammasome detection system is capable of identifying a large array of pathogen-connected molecular patterns, and it indicates that inflammatory caspases play a key role in protecting hosts from any type of pathogen (Man et al., 2017). For instance, DNA and RNA viruses can activate inflammasomes and trigger pyroptosis (Lupfer and Kanneganti, 2013; Man et al., 2016). However, studies have highlighted the harmful role played by pyroptosis in patients with HIV infection (Jakobsen et al., 2013; Monroe et al. 2014; Doitsh et al., 2014; Galloway et al., 2015; Munõz-Arias et al., 2015). Non-infected liver cells can also suffer caspase-1-dependent pyroptosis after witness cell infection with HCV (hepatitis C virus) (Kofahi et al., 2016). Studies conducted with mice have highlighted the important role played by caspase-1 and caspase-11 in infections caused by West Nile virus and influenza A virus (IAV) (Table 1) (Ramos et al., 2012).

Table 1. The role of caspase-1 and caspase-11 in response to viral infection in mice.

| Mouse | Virus | Phenotype compared to wild-type mice |
|---|---|---|
| Casp1 −/−Casp11 −/− | Encephalomyocarditis virus | No difference in survival (Rajan et al. 2011). |
| | Influenza A virus | Reduced survival (Thomas et al., 2012; Allen et al., 2009; Ichinohe et al., 2009), decreased IL-1β, IL-18, TNF, IL-6, KC, MIP-2 in the BALF, decreased neutrophils and monocytic dendritic cells in the BALF, diminished respiratory function (Thomas et al., 2009), decreased IFN-γ producing CD4+ and CD8+ T cells, reduced nasal IgA, increased pulmonary viral titer (Ichinohe et al., 2009). No difference in survival and body weight change in Mx1 sufficient host (Pillai et al., 2016) |
| | Murine gamma-herpesvirus 68 (MHV68) | No difference in viral burden in the lungs (Cieniewicz et al., 2015). |
| | Vesicular stomatitis virus | No difference in survival (Rajan et al., 2011). |
| | West Nile virus | Reduced survival (Ramos et al., 2012). |

## 3. Inflammasome

Zhao and Zhao (2020) have investigated PRRs (pattern recognition receptors), such as LRR, NACHT and PYD domains-containing protein- 3 inflammasome, which is a complex comprising the NOD (nucleotide-binding oligomerization) domain-like receptor NLRP3; this receptor is the adaptor of apoptosis-associated speck-like protein (ASC) containing a caspase-1 recruitment domain. The aforementioned complex plays a critical role in host's defense against microbes, since it stimulates IL-1β and IL-18 release and triggers pyroptosis, as previously mentioned. The NLRP3 is capable of detecting a diversity of PAMPs (pathogen-associated molecular patterns) and DAMPs (danger-associated molecular patterns) produced over viral replication; this replication process activates the NLRP3 inflammasome, which depends on antiviral immune responses to enable viral destruction. However, several viruses have developed sophisticated strategies, such as attacking the NLRP3 inflammasome to avoid the immune system (Sollberger et al., 2014). Few viruses are capable of inhibiting NLRP3 inflammasome stimulation to avoid innate immunity and enhance viral replication (Figure 1).

Viruses can suppress both the setting and the stimulation of NLRP3 inflammasomes on direct or indirect procedures. Proteins from Paramyxovirus V, measles virus and influenza virus NS1 can inhibit NLRP3 inflammasome induction and decrease IL-1β release. This interaction avoids NLRP3 self-oligomerization as the enrolment of ASC, and blocks inflammasome activation. The Epstein-Barr Virus (EBV) called miR-BART15 specifically attacks the miR223 binding site in the NLRP3 3′-UTR in order to suppress NLRP3 induction and transport non-infected cells, thus enhancing the immune-suppressive state. The virulent protein PB1-F2, codified by most IAV strains, can also impair NLRP3 inflammasome activation through different mechanisms. Protein degradation and NLRP3 ubiquitination is the regulatory mechanism solution for NLRP3 inflammasome induction. Several viral proteins can moderate NLRP3 degradation and inhibit NLRP3 inflammasome induction. These degradation processes can affect GSDMD and this anomalously breakage GSDMD derivative fails to activate infected-cell pyroptosis, which triggers viral replication and leads to viral escape. This interchange is partly indicative of the complex virus/host co-evolution and may enable potential treatments in the near future (Zhao and Zhao, 2020)

Besides microbial infections, type I interferon (I-IFN) signaling contributes to cell decease caused by IAV.

Type I-IFN signaling affects the positive regulation of ZBP1 innate immune sensor (known as DLM-1 or DAI (DNA-dependent activator of IFN-regulatory factors) (Kuriakose et al., 2016). ZBP1 recognizes the IAV nucleoprotein and the PB1 subunit of RNA polymerase; besides, it influences the activation of NLRP3 inflammasome, apoptosis and necroptosis through caspase-8, kinase RIPK3 and FADD (FAS-associated protein with death domain) (Kuriakose et al., 2016). These data were corroborated by other studies (Thapa et al., 2016; Nogusa et al., 2016), which highlighted the role played by the type I-IFN/ZBP1 axis in IV-induced cell death (Man et al., 2017a).

## 4. Mechanistic aspects

Some mechanistic aspects of pyroptosis in viruses were reported; for instance, IAV induces NLRP3 inflammasome (Kanneganti et al., 2006; Thomas et al., 2009; Allen et al., 2009; Kuriakose et al., 2016). Casp1 −/−Casp11 −/− mice are more susceptible to IAV infection than wild-type mice (Thomas et al., 2009; Allen et al., 2009; Ichinohe et al., 2009). They generate lesser IL-18 and IL-1β in the lungs, which reduce lung functions and increases viral titers (Thomas et al., 2009; Ichinohe et al., 2009). In fact, different congenic mice present different responses to infection caused by IAV. For instance, Casp1 −/−Casp11 −/− mice are as resistant as Mx1-positive congenic C57BL/6 mice to post-infection conditions; however, Casp1 and Casp11 deprivation decreases the mortality rate of Tlr7 −/−Mavs −/− mice (Pillai et al., 2016).

Although several studies focused on investigating the function of caspase-1 and caspase-11 has been conducted, their action *in vivo* remains poorly understood. IAVs and pathogens other than Gram-negative bacteria do not transport lipopolysaccharide (LPS), which is the PAMP inducing caspase-11. However, it is possible assuming that caspase-11 has the function of protecting hosts from microbial agents other than Gram-negative bacteria *in vivo*. The function of both caspase-1 and caspase-11 has been seen in host protection from *Aspergillus fumigatus* (fungal pathogen), which is a pathogen that does not transport LPS (Karki et al., 2015; Man et al., 2017b;). *A. fumigatus* induces NLRP3 inflammasome in THP-1 cells (human leukemia monocytic cell line), as well as AIM2 (absent in melanoma 2) and NLRP3 inflammasomes in mouse dendritic cells from bone-marrow and lung tissue (Saud-Sadier et al., 2010; Karki et al., 2015). Caspase-1-dependent response triggered the production of defensive cytokines IL-18 and IL-1β in a murine model of Aspergillosis (Karki etr al., 2015). Most importantly, Caspase-11 −/− mice are also highly sensitive to *A. fumigatus*-induced lethality in comparison to wild-type mice; however, collapse to infection with a late kinetics as compared with mice with the absence of caspase-1 or both caspase-11 and caspase-1 (Man et al., 2017a).

The caspase-11 activation or defense mechanism during *A. fumigatus* infections remains poorly understood. Caspase-11 activation might induce killing through actin-mediated phagosomal in order to control *A. fumigatus* propagation *in vivo* (Akhter et al., 2012; Caution et al., 2015; Walle et al., 2016). Caspase-1-dependent liberates IL-18 generates IFN-γ, which may supply a briefing signal for caspase-11 to regulate Aspergillosis *in vivo* (Karki et al., 2015). A study has suggested that, besides LPS, caspase-11 is capable of identifying oxidized phospholipids in hosts by means of an associated molecular pattern that triggers IL-1β release in dendritic cells without causing pyroptosis (Zanoni et al. 2016).

However, most recently, a study reported that caspase-1 and caspase-11 have triggered joint inflammation and pro-inflammatory cytokine generation in patients with brucellosis infection. Nevertheless, after a post-infection week, both caspases helped controlling Brucella joint infection by inducing pyroptosis and the IL-18 provided to the beginning of joint swelling and regulates Brucella joint infection. Thus, these data have indicated that inflammasomes trigger inflammation in a dependent manner through IL-18 and that pyroptosis restricts Brucella infection (Lacey et al., 2018).

Most recently, a study conducted with models such as cell lines, microbial strains, mice and human liver biopsy tissues, has found that pyroptosis effector protein gasdermin D (GSDMD) was broken by inflammasome-induced caspase-1 and LPS-activated caspase-11/4/5 . The fragmentation revealed domain where the pore were formed at gasdermin D-C terminal domain (GSDMD-C). Besides, the specific caspase-4/11 auto-processing site, which produces the p10 product, played an essential role in breaking GSDMD and in causing pyroptosis. The p10 product effectively binds to the GSDMD-C domain. The comparison between auto-processed and non-processed capase-11 structures has identified a β-sheet generated by auto-processing. The β-sheet assembles a hydrophobic binding GSDMD interface in the crystal structure of GSDMD-C/caspase-4/11 complex; that was only likely for p10 product from the complex. The crystalline arrangement of GSDMD-C/caspase-1 complex has shown similar GSDMD-identification mode (Wang et al., 2020a).

As previously mentioned, unlike apoptosis, pyroptosis comprises inflammation, the activation of a variety of caspases, mainly the caspase-1, which incises and polymerizes Gasdermin family members like GSDMD (Wang et al, 2019). The N-terminal cleavage product of GSDMD can cause extensive cell perforation by penetrating the lipid bilayer to induce high-order oligomerization in the membrane (Huang et al., 2017). This process is followed by cellular content release, which causes inflammation, cell swelling and lysis to prevent invading pathogens from replicating (Liu et al., 2016; Shi et al., 2017; Zhang et al., 2018). Once the cell is damaged (chemicals, disease-induced), several intracellular molecules with immunomodulatory activity are released from immune cells (Pandolfi et al., 2016). These molecules are called DAMPs, also known as alarm factor (Alcano et al., 2019). Similar to the aforementioned PAMPs, DAMPs can also be recognized by NLRP3 and lead to similar effects, such as pro-inflammatory responses and pyroptosis, as previously described (Ma et al., 2018; Gao et al., 2018).

Pyroptosis progresses faster than apoptosis and it is followed by the release of significant quantities of pro-inflammatory factors. Thus, unlike apoptosis, which is a neat, orderly-shrinking cell phagocytosis process, cell pyroptosis happens faster and triggers inflammation (Yang, 2020).

## 5. Coronavirus

Chen et al. (2019) have shown that SARS-CoV Viroporin 3a protein triggers NLRP3 inflammasome induction, as well as IL-1β release in bone marrow-derived macrophages, and it indicated that SARS-CoV caused cell pyroptosis. Jiang et al. (2019) observed that a MERS-CoV (HCoV-EMC/2012 strain) infected group presented increased expression of pyroptotic markers (NLRP3, IL-1β and caspase-1), as well as that pyroptosis can be suppressed by complement receptor C5aR1 (CD88).

It is known that once pathogens penetrate the cell, PAMPs are identified by PRRs, such as NLRP3, on the cell membrane and then bind to the precursor of caspase-1(pro-caspase-1) through the apoptosis-associated speck-like protein containing a caspase recruitment domain (ASC) in the cell in order to form a multiprotein complex, thus activating caspase-1 (He et al., 2016; Wang et al., 2019; Yang, 2020).

PRRs in the cells recognize signals triggered by bacteria and viruses, and bind to the precursor of caspase-1 through the adaptor protein ASC in order to form a multiprotein complex and activate caspase-1 (Wang et al., 2019). On the one hand, caspase1 incises gasdermin D to forms a peptide containing gasdermin D nitrogen-terminal active domain and causes extensive cell perforation by penetrating the lipid bilayer; this process also leads to cell rupture, cellular content release and triggers inflammatory reaction (Shi et al., 2017; Man et al., 2017). On the other hand, activated caspase-1 incises IL-1β and the precursor of IL-18 to form active IL-18 and IL-1β, which are released within the extracellular region in order to recruit inflammatory cells and spread the inflammatory reaction (Yang, 2020).

## 6. SARS-CoV-2 infection and cell pyroptosis

As it is known, 2019-nCoV virus was officially called SARS-CoV-2 and the disease was named COVID-19. A study review conducted by Yang et al. (2020) has summarized SARS history and epidemiology. Besides, other authors have pointed out several aspects, such as epidemiology, clinical features, pathogenesis, diagnosis and guidance of patients contaminated with SARS-CoV-2.

Epithelial and dendritic cells are activated at initial coronavirus contamination stages and form a cluster of pro-inflammatory cytokines and chemokines such as interleukins (IL-2, IL-6, IL-8, IL-1β), tumor necrosis factor (TNF), interferons (IFN-α/β), C-C motif chemokines and IP-10, among others. All these factors are regulated by the immune system. Therefore, excessive production of all these cytokines and chemokines enable the progress of the disease. Interleukin IL-10, which is generated by T-helper-2 (Th2), has antiviral activity; however, coronavirus infection can significantly inhibit it (Zhang et al., 2020a) (Fig.2).

Based on evidence available in the literature, patients contaminated with SARS-CoV-2 presented increased IL-1β (serum) secretion (Huang et al., 2020), which is a strong indicator of cell pyroptosis; it indicates that cell pyroptotic induction may be implicated in the pathogenesis of SARS-CoV-2-infected patients.

Although both non-classical and classical pyroptosis signaling are capable of activating IL-1β secretion, this process should be further investigated in SARS-CoV-2 pneumonia patients. Moreover, although both leucopenia and lymphopenia were observed in pneumonia patients, a larger number of SARS-CoV-2 contaminated patients had lymphopenia, which may indicate that lymphocytes are more likely to undergo cell pyroptosis during infection processes.

Chen et al. (2020a) observed the same behavior in 99 SARS-CoV-2-pneumonia patients assessed in their study. Almost all patients who died had lymphopenia. Other studies have also observed this finding in a significant number of patients, as follows: 138 hospitalized patients (Wang et al., 2020); 1,099 patients (Guan et al., 2020); 140 patients (Zhang et al., 2020b); 62 patients (Xu et al., 2020); 12 patients (Liu et al., 2020); 34 children (Wang et al., 2020b); 29 patients (Chen et al., 2020b); 138 patients (Wang et al., 2020c); 183 patients (Tang et al., 2020); 5,700 patients (Richarson et al., 2020); 21 patients (Arentz et al., 2020); and 9 pregnant women (Chen et al, 2020c). Interestingly, Chang et al. (2020) reported 13 patients with lymphopenia; all patients have recovered and presented marginally high levels of inflammatory markers, such as the number of lymphocytes. All these findings were corroborated by Lipi and Plebani (2020) and by Henry et al (2020).

It is known that either classical or non-classical pyroptosis can trigger IL-1β release, although the exact way it happens in patients with COVID-19 remains unknown. However, based on data available in the literature, SARS-CoV-2 likely causes cell pyroptosis, mainly in lymphocytes, due to NLRP3 inflammasome induction.

Then, some aspects still are necessary to study in this area, such as morphological changes of leucocytes and lymphocytes and classical and non-classical cell pyroptosis basic markers on biomolecules (nucleic acid, proteins). Although, in order to a rapid identification of SARS-CoV-2 cell pyroptosis in patients sample a rapid data such as NLRP3, IL-1beta, IL-18, and GSDMD are requested (Fig. 3) (Yan, 2020; Yang et al., 2000).

It is known that extensive cytokine release due to immune system reaction to viral contaminations can lead to cytokine storm and sepsis, which account for approximately 30% of COVID-19 death cases. This uncontrolled inflammation condition lies on multiple organ injuries and lead to organ failure, mainly in liver, cardiac and kidney functions. Patients contaminated with SARS-CoV who evolved to kidney failure ended up dying. On the other hand, cases in which recruited cells successfully healed lung infection, patients' immunological response retreated and they recovered. Other cases presented a dysfunctional immune reaction known as cytokine storm, which regulates extensive inflammation. COVID-19 patients in critical condition, who demanded rigorous care in hospitals, presented exacerbated parameters such as significantly high

10blood plasma levels of G-CSF (granulocyte colony-stimulating factor) and of interleukins, such as IL-2, 7 and 10, IP-10, MCP1 and MP1α (macrophage inflammatory protein 1α) and NTF (tumor necrosis factor); IL-6 cytokine levels increase in these situations and show higher concentrations in dead than in recovered patients. Most importantly, a significant inflammatory monocyte-derived FCN1+ macrophage population was observed in Broncho-alveolar material extracted from individuals with serous, although not mild, COVID-19. Besides, individuals at advanced disease stage have shown significant number of inflammatory monocytes (CD14+CD16+) in their peripheral blood than patients with mild COVID-19. These cellular systems release inflammatory cytokines that lead to cytokine storm by incorporating MIP1α, MCP1 and IP-10 (Tay et al., 2020, and references therein).

**Final remarks**

All studies about Covid-19 and, mainly, all aspects associated with pyroptosis point towards the need of conducting further studies focused on investigating morphological changes taking place in lymphocytes and leucocytes, as well as the expression of classical and non-classical cell pyroptosis markers at both nucleic acid and protein levels. It is important screening cell pyroptosis data such as NLRP3, IL-1beta, IL-18 and GSDMD in samples collected from SARS-CoV-2-pneumonia patients. Actually, very few reports on Covid-19 patients' follow-up address immunological aspects. The patient monitoring process is just based on the characterization of clinical aspects. However, this process must comprise the analysis of immunological data to help better understanding how this virus can be eliminated. The application of immunomodulation therapies to Covid-19 patients is the most promising treatment under investigation nowadays (Ingraham et al., 2020; Delafiori et al., 2020).

**Acknowledgments**
The authors would like to thank the São Paulo Research Council (FAPESP grant 2014/11154-1; 2018/10052-1), and the Brazilian National Council for Scientific and Technological Development (CNPq grant 552120/2011-1) Brazil.

**FIGURES**

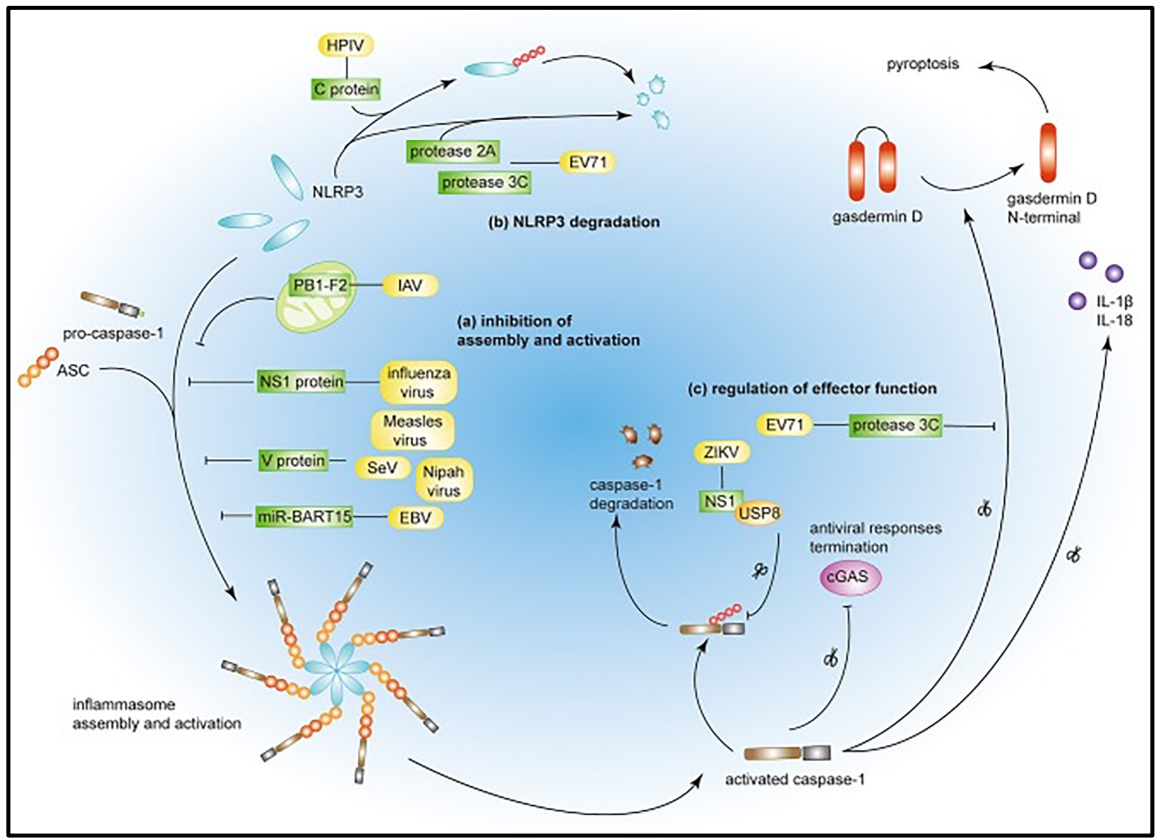

**Figure 1.** Viral immune evasion strategies by targeting the NLRP3 inflammasome. (a) Influenza virus NS1 protein, measles virus, SeV and Nipah virus V proteins prevent NRLP3 inflammasome assembly. PB1-F2 of IAV and miR-BART15 of EBV inhibit NLRP3 inflammasome activation. (b) EV71 proteases 2A and 3C and HPIV C protein induce NLRP3 protein degradation. (c) EV71 protease 3C and ZIKV NS1 protein modulate the effector function of the NLRP3 inflammasome by targeting GSDMD and caspase-1, respectively (Extracted from Zhao and Zhao, 2020, by permission of Frontiers Media S.A-Open Access, Creative Commons Attribution License).

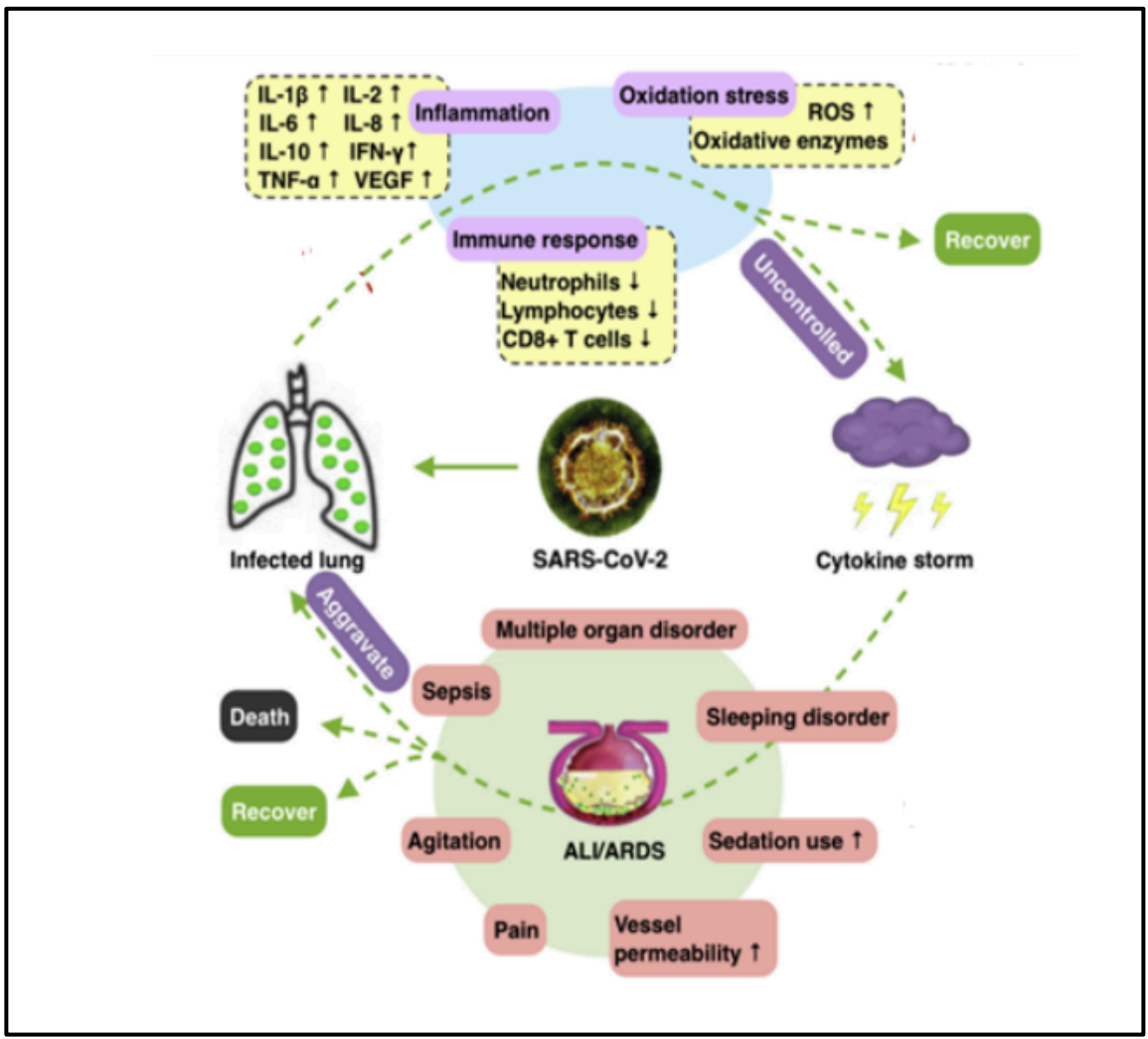

**Figure 2.** Pathogenesis of COVID-19. It is postulated that lungs infected by SARS-CoV-2, and a suppressed immune response, elevated inflammation and excessive oxidation stress proceed unabated, this results in the activation of the cytokine storm. Acute lung injury (ALI)/ Acute respiratory distress syndrome (ARDS) may ensue, accompanied by a series of complications, the outcomes of which vary according to the severity of the disease (Modified from Zhang et al., 2020a) (by permission from Elsevier).

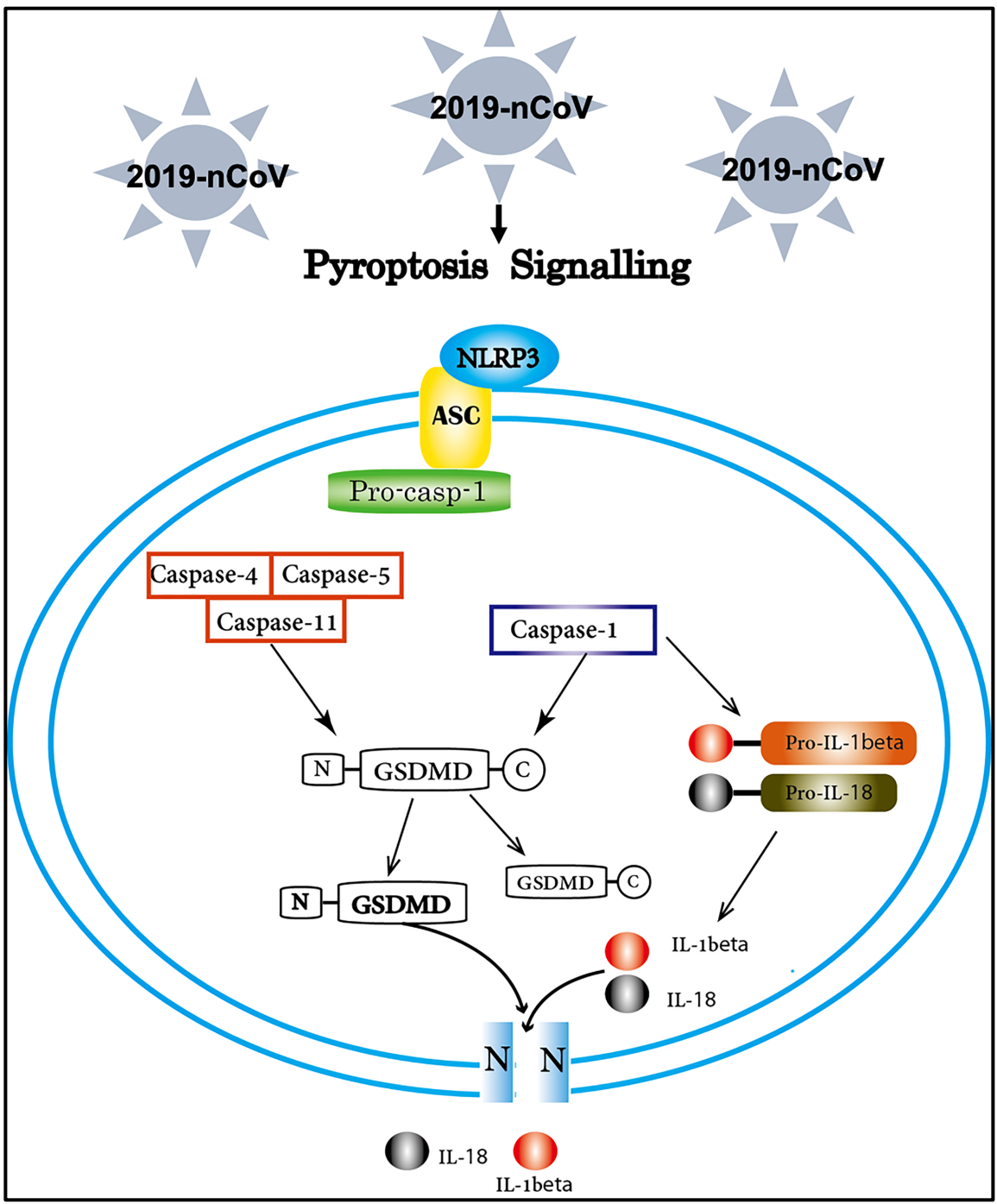

**Figure 3.** A hypothesis of the relationship between SARS-CoV-2 and cell pyroptosis. The COVID-19 may be linked to cell pyroptosis, especially in lymphocytes through the activation of the NLRP3 inflammasome. Morphological changes in lymphocytes and macrophages, nucleic acid and protein levels in classical and non-classical cells, detection of NLRP3 and GSDMD, and the role of inflammatory cytokines IL-1β and IL-18 requires further research (Electronic copy available at: https://ssrn.com/abstract=3527420).